\documentclass[a4paper,aps,prd,onecolumn,preprintnumbers,showpacs,nofootinbib]{revtex4}
\usepackage{amsmath,amssymb,graphics,epsfig,subfigure}
\usepackage{color}

\begin{document}
\newcommand {\nn} {\nonumber}
\renewcommand{\baselinestretch}{1.3}

\title{Black hole spectroscopy from null geodesics}

\author{Shao-Wen Wei \footnote{weishw@lzu.edu.cn},
        Yu-Xiao Liu \footnote{liuyx@lzu.edu.cn, corresponding author},
        and Chun-E Fu \footnote{fuche08@lzu.edu.cn}}

\affiliation{Institute of Theoretical Physics, Lanzhou University, Lanzhou 730000, People's Republic of China}

\begin{abstract}
The quasinormal mode frequencies can be understood from the massless particles trapped at the unstable circular null geodesics and slowly leaking out to infinity. Base on this viewpoint, in this paper, we construct the quantum entropy spectrum of the stationary black holes from the null geodesics using the new physical interpretation of the quasinormal mode frequencies proposed by Maggiore. Following this idea, we calculate the spacing of the entropy spectrum for various types of black holes with or without the charge and spin in any dimension $d$ of the spacetime. The result shows that the spacing closely depends on the charge, spin, and the dimension of the spacetime. Moreover, for a black hole far from the extremal case, the spacing is found to be larger than $2\pi\hbar$ for $d=4$, while smaller than $2\pi\hbar$ for $d\geq 5$, which is very different from the result of the previous work by using the usual quasinormal mode frequencies, where the spacing of the entropy spectrum is $2\pi\hbar$ and is independent of the black hole parameters and dimension $d$.
\end{abstract}



\maketitle

\section{Introduction}
\label{secIntroduction}

The quantization of the black hole horizon area and entropy has been a fascinating subject. The pioneer work can be traced back to Bekenstein \cite{Bekenstein1} with the famous conjecture that the black hole area should be represented by a quantum operator with a discrete spectrum of eigenvalue in a quantum gravity theory. Regarding the black hole horizon area as an adiabatic invariant, the equidistant area spectrum was obtained
\begin{eqnarray}
 A_{n}=\gamma \hbar\cdot n,
\end{eqnarray}
with $n$ an integer. From the dynamical modes of the classical theory, many attempts have been devoted to obtain such equally spaced area spectrum, whereas the spacing may be different from $\gamma=8\pi$ \cite{Bekenstein1995plb,Louko1996prd,Makela,
Dolgov1997plb,Peleg,Barvinsky2001plb,Barvinsky,Kastrup1996plb}.

In the respect that a black hole can be determined by a few parameters (such as mass, charge, and angular momentum), they come very close to our notion of elementary particles, especially like the hydrogen atom. Therefore characteristic vibrations of them, known as the quasinormal mode (QNM) frequencies \cite{Kokkotas,Nollert}, should play a significant role in black hole physics. The result is encouraging. Based on Bohr's correspondence principle that ``transition frequencies at large quantum numbers should
equal classical oscillation frequencies", Hod \cite{Hodprl1998,Hodprl19982} connected these classical oscillation frequencies with the real part of the highly damped QNM frequencies. Then he obtained the spacing of the area spectrum $\Delta A=(4\ln3)\hbar$ for a Schwarzschild black hole. On the other hand, Kunstatter \cite{Kunstatterprl2003} also derived the same area spectrum through the adiabatic invariant. The similar argument was also used to fix the Immirzi parameter by Dreyer \cite{Dreyerprl2003}. This rejuvenates a great interest in the investigation of the black hole area and entropy spectra via the interpretation of the QNM frequencies \cite{Polychronakosprd2004,Setarecqg2004,
Setareprd2004,Setareprd20042,Setareprd20043,Setareprd20044,Lepeplb2005}.

It has been, recently, pointed out by Maggiore \cite{Maggioreprl2008} that, in high damping limit, the proper frequency of the equivalent harmonic oscillator $\omega(E)$, which is interpreted as the QNM frequency, should be of the following form
\begin{eqnarray}
 \omega(E)=\sqrt{|\omega_{R}|^{2}+|\omega_{I}|^{2}},
\end{eqnarray}
with $\omega_{R}$ and $\omega_{I}$ the real and imaginary parts of the QNM
frequency, respectively. Following this idea, the Bekenstein's area spectrum will be recovered for a Schwarzschild black hole. Vagenas and Medved \cite{Vagenasjhep2008,Medvedcqg2008} applied this idea to derive the area spectrum of a rotating Kerr black hole with the choice of $\Delta\omega(E)=(|\omega_{I}|)_{n}-(|\omega_{I}|)_{n-1}$ for $\omega_{I}\gg\omega_{R}$. The area spectrum calculated with the modified Hod's method is equally spaced with spacing $\Delta A=8\pi\hbar$, which agrees with the Bekenstein's spacing. While employing the Kunstatter's method, the spectrum is nonequidistant. And the spacing is angular momentum dependent. Medved \cite{Medvedcqg2008} argued that the Kunstatter's method requires that the black hole should be far from the extremal one, i.e., small angular momentum limit. Considering this, the area spectra calculated from these two methods coincide with each other. Equivalently, the entropy spectrum can also be obtained using the Bekenstein-Hawking entropy/area law $S=A/4$, i.e., $S=2\pi\hbar\cdot n$ with spacing $\Delta S=2\pi\hbar$ in Einstein's gravity. However, when extended these results to the modified gravity theory, the area spectra were found not to be equidistant any more, while the entropy spectra were still in the form as that of the Einstein gravity \cite{Kothawalaprd2008,Weijhep2008}. Moreover, these results have been applied to different black holes \cite{Fernando,Lopez-Ortega,Kwon,WeiYang,KwonNam,Gonzalez,Lopez-Ortega2,ChenYang,KwonNam3,Li,LiuHu} and the same area and entropy spectra were observed. The spectra can also be found with the quantization of the angular momentum component \cite{Ropotenko,Jia}, which gives the same results as that obtained from the QNM frequencies. Black hole spectroscopy can also be obtained from the quantum tunneling method \cite{Banerjee10,Banerjee102,Jiang,MajhiVagenas11,Chen12,JiangHan,JiangHanCai,Ropotenko2,BanerjeeVagenas}.
In \cite{JiangHanCai}, the author argued that the entropy spectrum is $S_{n}=n\hbar$ with spacing $\Delta S=\hbar$ independent of the black hole parameters, gravity theory, and the dimension of the spacetime. The spacing was also suggested to be the lower bound of the entropy spectrum \cite{BanerjeeVagenas}.

As we know, the QNM frequencies are determined by solving the perturbation equation with the boundary conditions that only purely outgoing at infinity and ingoing at the horizon. In a black hole background, null geodesics appear very useful to explain the QNM frequencies \cite{Press,Goebel,Ferrari,Mashhoon,Berti,Cardoso}. The QNM frequencies can be interpreted in terms of the massless particles trapped at the unstable circular null geodesics and slowly leaking out to infinity. The real part of the QNM frequencies corresponds to the angular velocity at the unstable null geodesics, and the imaginary part is measured by the instability time scale of the orbit. In the eikonal limit ($m\gg1$), the QNM frequencies of the black holes approximately read
\begin{eqnarray}
 \omega_{\text{QNM}}=\Omega_{c}m-i(n+1/2)|\lambda|,\label{QNM}
\end{eqnarray}
with $\Omega_{c}$ the angular velocity at the unstable circular null
geodesic and $\lambda$ Lyapunov exponent. This result is not only valid for a static, spherically symmetric and asymptotically flat line element in any dimension, but also for the equatorial orbits in the geometry of a rotating Myers-Perry black hole in higher dimensions \cite{Cardoso}. The parameters $\Omega_{c}$ and $\lambda$, as we will show, can be both expressed in terms of the radius of the unstable circular null geodesics for an arbitrary stationary black hole with the asymptotics (\ref{asymptotics}).

Based on such view of the QNM frequencies, the strong gravitational black hole lensing, high-energy absorption cross section were found to be related to the unstable circular null geodesics \cite{Stefanov,WeiLiu} with some impressive results obtained. It is therefore imperative to find the relation between the thermodynamic quantity of a black hole and the unstable circular null geodesics. Motivated by this idea, in this paper, we aim to study the entropy spectrum of the black holes from such view of the QNM frequencies. Employing the Hod's method, we found that the entropy spectrum of an asymptotically flat black hole can be expressed in terms of its unstable circular null geodesics. And the spacing of the entropy spectrum is found to depend on the charge, spin, and the dimension of the spacetime, which is very different from that of the previous work by using the usual QNM frequencies \footnote{In order to distinguish the QNM frequencies obtained through solving the perturbation equation from such view, we refer these from the perturbation equation as the usual QNM frequencies.}. Moreover, the result shows that the spacing $\Delta S$ of the entropy spectrum decreases with the dimension $d$ of the spacetime. For $d=4$, $\Delta S$ is larger than $2\pi\hbar$, while for $d\geq 5$, it is smaller than $2\pi\hbar$. Moreover, the spacing $\Delta S$ will be below the low bound obtained from the tunneling method as suggested in \cite{BanerjeeVagenas} when $d$ is larger than 151.

The paper is organized as follows. In section \ref{geodesic}, we investigate the null geodesics of an asymptotically flat spacetime. Then we get the relationship between the entropy spectrum and the null geodesics. In section \ref{Static}, we explore the entropy spectrum in a static and spherically symmetric black hole. And we generalize it to the stationary and axis-symmetric black hole in section \ref{Stationary}. The final section is
devoted to a brief summary.

\section{Entropy spectroscopy and null geodesics}
\label{geodesic}

In this section, we first study the null geodesics of a black hole, and then show the QNM frequencies obtained from the view described above. Further, based on such view, we interpret the black hole entropy spectrum through the null geodesics.

Here we assume that the equatorial metric ($\theta=\pi/2$) of a black hole background is in the following simple form
\begin{eqnarray}
 ds^{2}=-A(r)dt^{2}+B(r)dr^{2}+C(r)d\phi^{2}-D(r)dtd\phi,\label{metric0}
\end{eqnarray}
which can describe the equatorial plane of a static, spherically symmetric black hole or a stationary, axis-symmetric black hole. We only require that these metric functions satisfy the following proper asymptotics
\begin{eqnarray}
 A(r\rightarrow \infty)=1,\;\;B(r\rightarrow \infty)=1,\;\;
 C(r\rightarrow \infty)=r^{2},\;\;D(r\rightarrow \infty)=0. \label{asymptotics}
\end{eqnarray}
The geodesics of a massless particle in the equatorial plane of such black hole background (\ref{metric0}) can be easily obtained via the Lagrangian, which reads
\begin{eqnarray}
 2\mathcal{L}=g_{\mu\nu}\dot{x}^{\mu}\dot{x}^{\nu}=-A(r)\dot{t}^{2}
              +B(r)\dot{r}^{2}+C(r)\dot{\phi}^{2}
              -D(r)\dot{t}\dot{\phi}.
\end{eqnarray}
The dot over a symbol denotes the ordinary differentation with respect to an affine parameter. The generalized momentum reproduced from this Lagrangian is $p_{\mu}=\frac{\partial \mathcal{L}}{\partial \dot{x}^{\mu}}=g_{\mu\nu}\dot{x}^{\nu}$ with its components given by
\begin{eqnarray}
 p_{t}&=&-A(r)\dot{t}-\frac{D(r)}{2}\dot{\phi}\equiv-E,\label{PT}\\
 p_{\phi}&=&-\frac{D(r)}{2}+C(r)\dot{\phi}\equiv l,\label{Pphi}\\
 p_{r}&=&B(r)\dot{r}.
\end{eqnarray}
The parameter $E$ is the energy of the particle and $l$ is the orbital angular momentum of the particle in the $\phi$ direction measuring by an observer at
rest at infinity. Solving (\ref{PT}) and (\ref{Pphi}), we have the $t$-motion and $\phi$-motion
\begin{eqnarray}
 \dot{t}=\frac{4C(r)E-2D(r)l}{4A(r)C(r)+D(r)^{2}},\quad \dot{\phi}=\frac{2D(r)E+4A(r)l}{4A(r)C(r)+D(r)^{2}}.\label{tphi0}
\end{eqnarray}
The Hamiltonian of this system is
\begin{eqnarray}
 2\mathcal{H}&=&2(p_{\mu}\dot{x}^{\mu}-\mathcal{L}) \nonumber\\
             &=&-A(r)\dot{t}^{2}+B(r)\dot{r}^{2}
              +C(r)\dot{\phi}^{2}-D(r)\dot{t}\dot{\phi}=\delta.\label{HH}
\end{eqnarray}
Here $\delta=-1, 0, 1$ are for timelike, null, and spacelike geodesics, respectively. Since we focus on the null geodesics, we take $\delta=0$. Then inserting (\ref{tphi0}) into (\ref{HH}), we get the $r$-motion
\begin{eqnarray}
 \dot{r}^{2}=\frac{4}{B(r)}\left(\frac{C(r)E^{2}-D(r)El-A(r)l^{2}}{4A(r)C(r)+D(r)^{2}}\right),
\end{eqnarray}
which can be further rewritten as
\begin{eqnarray}
 \dot{r}^{2}+V_{\text{eff}}=0,
\end{eqnarray}
with the effective potential $V_{\text{eff}}=-\frac{4}{B(r)}\left(\frac{C(r)E^{2}-D(r)El-Al^{2}}
{4A(r)C(r)+D(r)^{2}}\right)$. The unstable circular orbit is determined by $V_{\text{eff}}$ through the following conditions:
\begin{eqnarray}
 V_{\text{eff}}=0,\quad \frac{\partial V_{\text{eff}}}{\partial r}=0, \quad
 \frac{\partial^{2} V_{\text{eff}}}{\partial r^{2}}<0.
\end{eqnarray}
The third condition ensures the instability of the orbit. And from the above conditions, it is easy to find that the unstable circular orbit is located at the local maxima of the effective potential. And the first two conditions yield
\begin{eqnarray}
 &&A(r)\tilde{l}^{2}+D(r)\tilde{l}-C(r)=0, \label{LL}\\
 &&A'(r)\tilde{l}^{2}+D'(r)\tilde{l}-C'(r)=0,\label{LL2}
\end{eqnarray}
where the impact parameter $\tilde{l}=l/E$, and the prime denotes the derivative with respect to $r$. Solving (\ref{LL}), we get the impact parameter
\begin{eqnarray}
 \tilde{l}=\frac{-D(r)+\sqrt{4A(r)C(r)+D(r)^{2}}}{2A(r)}.
\end{eqnarray}
The minimum impact parameter $\tilde{l}_{c}$ is measured at $r=r_{c}$ with $r_{c}$ the radius of the unstable circular orbit, which satisfies the equation derived from (\ref{LL2})
\begin{eqnarray}
 A_{c}C'_{c}-A'_{c}C_{c}+\tilde{l}_{c}(A'_{c}D_{c}-A_{c}D'_{c})=0,\label{Rcir}
\end{eqnarray}
where the down index ``c" means that these functions are evaluated at $r_{c}$. Given the explicit forms of the metric functions, we can obtain the radius of the unstable circular orbit, which corresponds to the largest root of equation (\ref{Rcir}) with the unstable condition satisfied. For the Schwarzschild black hole, we easily get $r_{c}=3M$. The Lyapunov exponent $\lambda$ and angular velocity $\Omega_{c}$ are two important quantities to measure the property of the unstable circular null geodesics, which are defined as
\begin{eqnarray}
 \lambda=\sqrt{\frac{V_{\text{eff}}''}{2\dot{t}^{2}}}\bigg|_{r_{c}},\quad
 \Omega_{c}=\frac{\dot{\phi}}{\dot{t}}\bigg|_{r_{c}}.
\end{eqnarray}
Combining by the effective potential and the null geodesics, we find that
\begin{eqnarray}
 \lambda=\frac{\kappa}{\tilde{l}_{c}}, \quad
 \Omega_{c}=\frac{1}{\tilde{l}_{c}},
\end{eqnarray}
with $\kappa$ in a compacted form
\begin{eqnarray}
 \kappa^{2}=\frac{A_{c}C''_{c}-A''_{c}C_{c}+\tilde{l}_{c}(A''_{c}D_{c}-A_{c}D''_{c})}
             {2A_{c}B_{c}}.
\end{eqnarray}
In \cite{Cardoso}, the author showed that, for the $d$-dimensional Myers-Perry black holes, the real part of the QNM frequency, or the angular frequency of the equatorial circular geodesics is the inverse of their impact parameter. Here we clearly show that this is a general property and it holds for any stationary, asymptotically flat black hole. Inserting the parameters $\lambda$ and $\Omega_{c}$ into (\ref{QNM}), we will get the QNM frequency corresponding to the metric (\ref{metric0}),
\begin{eqnarray}
 \omega=\frac{1}{\tilde{l}_{c}}(m-i(n+1/2)\kappa).
\end{eqnarray}
Then the vibrational frequency $\Delta \omega$ could be written as
\begin{eqnarray}
 \Delta \omega&=&\sqrt{(\Delta\omega_{R})^{2}+(\Delta\omega_{I})^{2}}\nonumber\\
       &=&\frac{1}{\tilde{l}_{c}}\sqrt{1+\kappa^{2}},\label{varomega}
\end{eqnarray}
where $\Delta\omega_{R}=(|\omega_{R}|)_{m}-(|\omega_{R}|)_{m-1}$ and  $\Delta\omega_{I}=(|\omega_{I}|)_{n}-(|\omega_{I}|)_{n-1}$.

According to the Bohr-Sommerfeld quantization
rule, substituting this vibrational frequency into the Clausius relation $\Delta Q=\hbar \Delta\omega=T\Delta S$ yields the entropy spectrum
\begin{eqnarray}
 S=\frac{\hbar \Delta\omega}{T}\cdot n=\frac{\sqrt{1+\kappa^{2}}}{T \tilde{l}_{c}}\hbar\cdot n,
\end{eqnarray}
with spacing
\begin{eqnarray}
 \Delta S=\frac{\sqrt{1+\kappa^{2}}}{T \tilde{l}_{c}}\hbar,
\end{eqnarray}
where $T$ is the Hawking temperature of the black hole. From this expression, we see that the spacing depends both on the temperature of the black hole and the property of the unstable circular null geodesics. Using the Bekenstein-Hawking entropy/area law $S=A/4$, the area spectrum can also be obtained, which is also obviously dependent of the temperature and the null geodesics.

To close this section, we would like to remark that, the area and entropy spectra obtained from such view of the QNM frequencies is not only valid for a static and spherically symmetric black hole, but also for a stationary and axis-symmetric black hole with the asymptotics (\ref{asymptotics}). In order to work out more details about the spectra, we will apply this method to different black holes in the next of this paper.

\section{Static and spherically symmetric black holes}
\label{Static}

In this section, we would like to apply the above method to study the entropy spectrum of the static and spherically symmetric black holes in any dimension.

\subsection{$d$-dimensional Schwarzschild black holes}

Here, let us consider a specific example, the $d$-dimensional Schwarzschild-Tangherlini metric, which is
\begin{eqnarray}
 &&ds^{2}=-f(r)dt^{2}+f^{-1}(r)+r^{2}d\Omega^{2}_{(d-2)},\label{Sch}\\
 &&f(r)=1-\bigg(\frac{r_{h}}{r}\bigg)^{d-3},
\end{eqnarray}
where $d\Omega^{2}_{(d-2)}$ is the line element of the unit
$(d-2)$-dimensional sphere $S^{(d-2)}$ with the usual
angular coordinates $\phi \in [0,\;2\pi]$ and $\theta_{i} \in [0,\;\pi]$ ($i=1,2,...,d-3$). The parameter $r_{h}$ denotes the outer horizon radius of the black hole, which is related to the ADM mass of the spacetime as $M=\frac{(d-2)A_{(d-2)}r_{h}^{d-3}}{16\pi}$ with $A_{(d-2)}$ the area of a unite $(d-2)$-dimensional sphere. The temperature of this black hole is calculated as
\begin{eqnarray}
 T^{\text{sch}}=\frac{\partial_{r}f(r)}{4\pi}\bigg|_{r_{h}}
        =\frac{(d-3)}{4\pi r_{h}}.
\end{eqnarray}
Note that $T^{\text{sch}}$ is proportional to the inverse of the horizon radius, which means a supermassive black hole has a low temperature, while it has strong gravitational effect. This result holds for any dimension of the spacetime.

Next, let us examine the null geodesics of this spacetime. We only restrict our attention to the equatorial hyperplane defined by $\theta_{i}=\pi/2$ with $i=1,..., (d-3)$. Then the metric reduces to
\begin{eqnarray}
 ds^{2}=-f(r)dt^{2}+f^{-1}(r)dr^{2}+r^{2}d\phi^{2},
\end{eqnarray}
Comparing with (\ref{metric0}), we easily get
\begin{eqnarray}
 A=f(r),\;B=f^{-1}(r),\; C=r^{2},\; D=0.\label{RNreduce}
\end{eqnarray}
Thus, by solving equation (\ref{Rcir}), the radius of the unstable circular geodesics will be obtained
\begin{eqnarray}
 r_{c}=\bigg(\frac{(d-1)}{2}\bigg)^{1/(d-3)}r_{h}.
\end{eqnarray}
It is worth noting that we always have $r_{c}>r_{h}$ for any value of $d\geq 4$. When $d=4$, we get the usual result that $r_{c}=\frac{3}{2}r_{h}$. And when $d\rightarrow \infty$, the unstable circular orbits radius will approach the horizon. A straightforward calculation shows
\begin{eqnarray}
 \tilde{l}_{c}&=&\sqrt{\frac{2}{(d-3)}}\bigg(\frac{d-1}{2}\bigg)^{\frac{(d-1)}{2(d-3)}}r_{h},\\
 \kappa&=&\sqrt{d-3}.
\end{eqnarray}
Thus, the vibrational frequency deduced from (\ref{varomega}) reads
\begin{eqnarray}
 \Delta \omega=(d-2)\bigg(1-\big(2(d-1)\big)^{3-d}\bigg)\frac{1}{r_{h}}.
\end{eqnarray}
Then, the spacing of the entropy spectrum is
\begin{eqnarray}
 \Delta S=2^{\frac{2d-5}{d-3}}\pi\bigg(d-1\bigg)^{\frac{d-1}{2(3-d)}}
       \sqrt{\frac{d-2}{d-3}}\hbar.
\end{eqnarray}
From this equation, it is clear that the spacing of the entropy spectrum depend on the dimension $d$ of the spacetime. And when $d\rightarrow \infty$, we find the spacing monotonically decreases to zero. We list the spacing of the entropy spectrum for $d=4-10$ in table \ref{Table1}. One easily learns from the table that, for $d=4$, the spacing is the largest one with $\Delta S\approx 2.1774\pi\hbar>2\pi\hbar$. However this spacing will be smaller than  $2\pi\hbar$ when $d\geq 5$. Therefore this result is very different from these obtained in the previous work from the usual QNM frequencies, where the spacing is $2\pi\hbar$ and independent of the dimension $d$. Moreover, we are aware that the spacing $\Delta S$ will be below the low bound obtained from the tunneling method as suggested in \cite{BanerjeeVagenas} for $d\geq 151$.

\begin{table}[h]
\begin{center}
\begin{tabular}{|c|c|c|c|c|c|c|c|}
  \hline
   $d$& 4& 5& 6& 7& 8& 9& 10\\
\hline
   $\frac{\Delta S}{2\pi\hbar}$
   & 1.0887 & 0.8660 & 0.7610 & 0.6936 & 0.6446 & 0.6062 & 0.5749 \\
\hline
\end{tabular}
\caption{The spacing of the black hole entropy spectrum for different values of the dimension $d$ of the spacetime.}\label{Table1}
\end{center}
\end{table}

\subsection{$d$-dimensional Reissner-Nordstr$\ddot{o}$m black holes}

It is natural to speculate that the presence of the charge will dramatically change the spacing of the entropy spectrum for a Schwarzschild black hole. So it is worth to examine, the charged Reissner-Nordstr$\ddot{o}$m black hole in any dimension.

\begin{figure}
\centerline{\includegraphics[width=8cm]{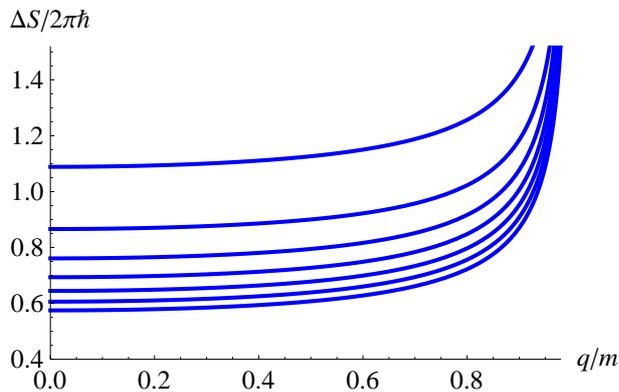}}
\caption{The spacing of the entropy spectrum for the Reissner-Nordstr$\ddot{o}$m black holes as a function of the dimensionless charge parameter $q/m$ for $d=4,5,6,7,8,9,10$ from top to bottom.}\label{PRN}
\end{figure}

The line element describing this spacetime is in the same form as (\ref{Sch}), while with different metric function
\begin{eqnarray}
 f(r)=1-\frac{2m}{r^{d-3}}+\frac{q^{2}}{r^{2(d-3)}},\label{RN}
\end{eqnarray}
where the parameters $m$ and $q$ are linked to the mass $M$ and charge $Q$ as
\begin{eqnarray}
 m=\frac{8\pi M}{(d-2)A_{d-2}}, \quad
 q^{2}=\frac{8\pi Q^{2}}{(d-2)(d-3)A_{d-2}}.
\end{eqnarray}
Solving $f(r)=0$, we get the horizons located at
\begin{eqnarray}
 r_{\pm}^{d-3}=m\pm\sqrt{m^{2}-q^{2}}.
\end{eqnarray}
From this equation, we note that there may exist two horizons for $q/m<1$, one horizon for $q/m=1$, or no horizon for $q/m>1$. Here we are only interesting in the noextremal black hole with outer horizon $r_{h}=r_{+}$. Therefore, the temperature of the outer horizon is
\begin{eqnarray}
 T^{\text{RN}}=\frac{(d-3)}{2\pi}r_{h}^{2-d}(m-r_{h}^{3-d}q^{2}).
\end{eqnarray}
In the equatorial hyperplane, the reduced metric has the form of (\ref{RNreduce}) with $f(r)$ given by (\ref{RN}).
The radius of the unstable circular orbit is calculated as
\begin{eqnarray}
 r_{c}^{d-3}=\frac{(d-1)m+\sqrt{(d-1)^{2}m^{2}-4(d-2)q^{2}}}{2}.
\end{eqnarray}
For $d=4$, we have $r_{c}=\frac{3M+\sqrt{9M^{2}-8Q^{2}}}{2}$, which implies that in the range $1<\frac{Q^{2}}{M^{2}}\leq 9/8$, the naked singularity located at $r=0$ is surrounded by a photon sphere rather than a horizon. A simple calculation shows
\begin{eqnarray}
 \tilde{l}_{c}&=&\frac{r_{c}^{d-2}}{\sqrt{r_{c}^{2(d-3)}-2mr_{c}^{d-3}+q^{2}}},\\
 \kappa&=&r_{c}^{3-d}\sqrt{r_{c}^{2(d-3)}+(d-1)(d-4)mr_{c}^{d-3}-(d-2)(2d-7)q^{2}}.
\end{eqnarray}
Then the spacing of the entropy spectrum is
\begin{eqnarray}
 \Delta S=&&\frac{2\pi r_{h}^{d-2} \sqrt{1-2r_{c}^{(3-d)}m+r_{c}^{2(3-d)}q^{2}}}{(d-3)r_{c}(m-r_{h}^{3-d}q^{2})}
  \nonumber\\
  &&\times\sqrt{2-(d-2)(2d-7)r_{c}^{2(3-d)}q^{2}+(d-4)(d-1)r_{c}^{3-d}m}\hbar.
\end{eqnarray}
For a clear view, we plot this spacing in Figure \ref{PRN} for different values of $d$ and the dimensionless charge parameter $q/m$. When $q=0$, the result will reduce to the Schwarzschild black hole case. Although the spacing depends on the charge $q/m$, however from the figure, we find that the spacing for the same $d$ varies very slowly in the small charge case. When the dimensionless charge parameter $q/m$ approaches one, the black hole will approach an extremal one, and the spacing blows up mainly due to the vanishing temperature of the extremal black hole.

\section{Stationary and axis-symmetric black holes}
\label{Stationary}

In the above section, we consider the static, spherically symmetric black holes. The result shows that the spacing of the entropy spectrum depends on the dimension $d$ and dimensionless charge $q/m$ of the black holes. And the spacing is larger than $2\pi\hbar$ for $d=4$, and smaller than $2\pi\hbar$ for $d\geq 5$ in small charge, which is very different from the result obtained through the usual QNM frequencies. In this section, we will generalize this method to the stationary and axis-symmetric black holes: four-dimensional Kerr black holes and $d$-dimensional Myers-Perry black holes.

\subsection{Kerr black holes}

\begin{figure}
\centerline{\includegraphics[width=8cm]{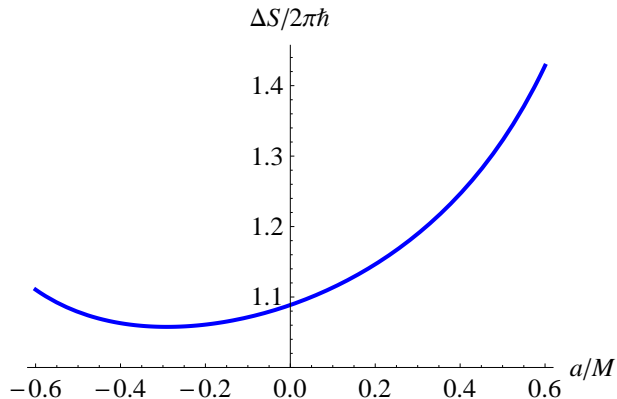}}
\caption{Behavior of the spacing of the entropy spectrum for the Kerr black hole.}\label{Pkerr}
\end{figure}

In Boyer-Lindquist coordinates, the four-dimensional Kerr metric is expressed as
\begin{eqnarray}
 ds^{2}=&-&\bigg(1-\frac{2Mr}{\rho^{2}}\bigg)dt^{2}+\frac{\rho^{2}}{\Delta}dr^{2}+\rho^{2}d\theta^{2}
        +\bigg(a^{2}+r^{2}+\frac{2a^{2}Mr\sin^{2}\theta}{\rho^{2}}\bigg)\sin^{2}\theta d\phi^{2}\nonumber\\
        &-&\frac{4aMr\sin^{2}\theta}{\rho^{2}}dtd\phi,
\end{eqnarray}
where $\Delta=r^{2}-2Mr+a^{2}$ and $\rho^{2}=r^{2}+a^{2}\cos^{2}\theta$. Here $M$ and $a$ are the mass and spin of the black hole, respectively. It is well known that this black hole has two horizons determined by $\Delta=0$, i.e., $r_{\pm}=M\pm\sqrt{M^{2}-a^{2}}$. When $a/M<1$, there is the outer horizon $r_{h}=r_{+}$ endowed with a temperature
\begin{eqnarray}
 T^{\text{Kerr}}=\frac{\sqrt{M^{2}-a^{2}}}{4\pi M(M+\sqrt{M^{2}-a^{2}})}.
\end{eqnarray}
In the equatorial plane, the metric is of the general form (\ref{metric0}) with metric functions given by
\begin{eqnarray}
 A(r)=1-\frac{2M}{r},\quad B(r)=\frac{r^{2}}{\Delta},\quad
 C(r)=r^{2},\quad D(r)=\frac{4aM}{r}.
\end{eqnarray}
Note that when the spin $a$ approaches 0, the Kerr black hole will reduce to a Schwarzschild black hole. And in the equatorial plane, we have a vanishing $D(r)$. Solving (\ref{Rcir}), we get the circular orbit radius
\begin{eqnarray}
 r_{c}=2M\bigg(1+\cos\bigg(\frac{2}{3}\arccos(\mp|a|/M)\bigg)\bigg).
\end{eqnarray}
Here, the upper sign is for the corotating orbit, while the lower sign is for the counterrotating orbit. It is also worth noting that the counterrotating orbit is located far away from the horizon than the corotating orbit. Then we are allowed to calculate the parameters $\tilde{l}_{c}$ and $\kappa$,
\begin{eqnarray}
 \tilde{l}_{c}&=&\frac{2aM-r_{c}\sqrt{\Delta}_{c}}{2M-r_{c}},\\
 \kappa&=&\sqrt{\frac{\Delta_{c}[r^{2}_{c}(r_{c}-2M)+4aM(a-\sqrt{\Delta_{c}})]}
      {r_{c}^{3}(r_{c}-2M)^{2}}},
\end{eqnarray}
with $\Delta_{c}=\Delta(r_{c})$. Then we get the spacing of the entropy spectrum
\begin{eqnarray}
 \Delta S=&&\frac{\sqrt{\Delta_c \left(4 a^2 M+r_c^2 \left(r_c-2M\right)\right)
   -4 a M \Delta _c^{3/2}+r_c^3 \left(r_c-2 M\right)^2}}
    {r_c^{3/2}\sqrt{M^2-a^2}
   \left(r_c \sqrt{\Delta_c}-2aM\right)} \nonumber\\
    &&\times 4\pi M \left(M+\sqrt{M^2-a^2}\right)\hbar.
\end{eqnarray}
The behavior of this spacing is shown in Figure \ref{Pkerr}. If we restrict our attention to the far from extremal black hole, i.e., in the neighborhood of $a/M=0$, one easily obtains a spacing larger than $2\pi\hbar$. And when $a/M=0$, it reduces to the Schwarzschild black hole case in $d=4$. On the other hand, one can find that for a Kerr black hole with spin $|a|$, there exist two unstable circular null geodesics, the counterrotating orbit and corotating orbit, which implies that there are two sets of the QNM frequencies, and two different entropy spectra will be obtained. Thus the entropy spectrum obtained in this way is seems to depend on the sign of the orbit angular momentum of the massless particle. However, it is not the case. In \cite{Zimmerman}, the authors clearly showed that there exist two distinct sets of the QNM frequencies for nearly extremal Kerr black holes. Here we can conjecture that this result is also held for an arbitrary rotating black hole. And these two distinct sets could be interpreted with the counterrotating and corotating orbits, respectively. So the entropy spectra obtained from the counterrotating and corotating orbits are in fact calculated with the two distinct sets of QNM frequencies. Therefore the spectra have no relation with the sign of the orbit angular momentum. From Figure \ref{Pkerr}, one easily finds that the spacing related to the counterrotating orbit has the smaller value than the corotating one, which can be understood as that the spacing related to the counterrotating orbit may more approach the minimum spacing of the black hole entropy.

\subsection{$d$-dimensional Myers-Perry black holes}

In this subsection, we would like to generalize this method to these black holes with spin in higher dimensions.

\begin{figure}
\centerline{\includegraphics[width=8cm]{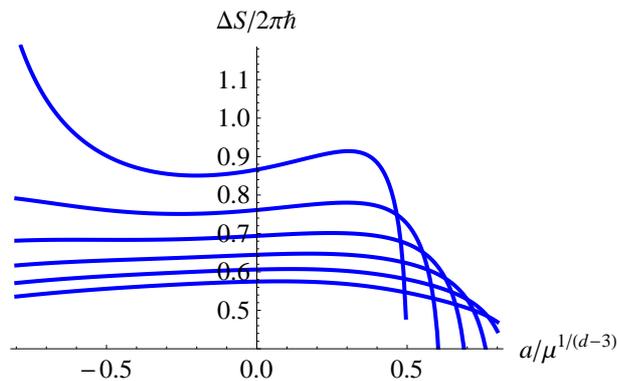}}
\caption{The spacing of the entropy spectrum for the Myers-Perry black holes as a function of the dimensionless spin parameter $a/\mu^{1/(d-3)}$ for $d=5,6,7,8,9,10$ from top to bottom.}\label{PMP}
\end{figure}

In \cite{Myers}, Myers and Perry found a general black hole solution in asymptotically flat spacetime in $d$ dimensions with all $\frac{(d-1)}{2}$ $a_{i}$ nonvanishing. Here, we only consider the simple case that only a non-zero spin $a_{1}=a\neq 0$, therefore the metric can be written as
\begin{eqnarray}
 ds^{2}=&-&dt^{2}+\frac{\mu}{r^{d-5}\rho^{2}}(dt+a\sin^{2}\theta d\phi)^{2}
        +\frac{\rho^{2}}{\Delta}dr^{2}+\rho^{2}d\theta^{2}\nonumber\\
        &+&(r^{2}+a^{2})\sin^{2}\theta d\phi^{2}
        +r^{2}\cos^{2}\theta d\Omega^{2}_{(d-4)},
\end{eqnarray}
with $\rho^{2}=r^{2}+a^{2}\cos^{2}\theta$ and $\Delta=r^{2}+a^{2}-\frac{\mu}{r^{d-5}}$. The parameters $\mu$ and $a$ are related to the black hole mass and angular momentum in the following forms
\begin{eqnarray}
 M=\frac{(d-2)\Omega_{d-2}}{16\pi G}\mu,\quad
 J=\frac{2}{d-2}Ma.
\end{eqnarray}
The outer horizon is determined as the largest root of $\Delta(r)=0$, which gives
\begin{eqnarray}
 r_{h}^{2}+a^{2}-\frac{\mu}{r_{h}^{d-5}}=0.
\end{eqnarray}
For $d=4$, it reduces to the Kerr black hole with $r_{h}=M+\sqrt{M^{2}-a^{2}}$. And for $d=5$, we easily get $r_{h}=\sqrt{\mu-a^{2}}$. Thus these two black hole cases are bounded with a maximum spin $a_{max}$. While when $d\geq 6$, there is an interesting result that the spin $a$ of the black hole could be arbitrary large, which is referred as the ``ultra-spinning" black hole. However, here we only consider the small $a$ limit. The black hole temperature is calculated as
\begin{eqnarray}
 T^{\text{MP}}=\frac{1}{4\pi}(\frac{2r_{h}^{d-4}}{\mu}+\frac{d-5}{r_{h}}).
\end{eqnarray}
In the equatorial hyperplane, the metric is of the general form (\ref{metric0}) with metric functions given by
\begin{eqnarray}
 A(r)=1-r^{3-d}\mu,\quad B(r)=\frac{r^{2}}{\Delta},\quad
 C(r)=r^{2}+a^{2}(1+r^{3-d}\mu),\quad D(r)=\frac{2a\mu}{r^{d-3}}.
\end{eqnarray}
The radius of the circular orbit is determined by (\ref{Rcir}), which leads to
\begin{eqnarray}
 2 r^{2d-6}-(d+1)\mu r^{d-3}+2a\mu(d-3)(\sqrt{\Delta}-a)r^{d-5}+(d-1)\mu^{2}=0.
\end{eqnarray}
Generally, this equation cannot be solved analytically. However for small $d$, the exact result can be obtained. For $d=4$, it reduces to the Kerr case. And for $d=5$, we get $r_{c}=\sqrt{2}\sqrt{\mu\pm a\sqrt{\mu}}$ for counterrotating and corotating orbits, respectively. The parameters $\tilde{l}_{c}$ and $\kappa$ are
\begin{eqnarray}
 \tilde{l}_{c}&=&\frac{-a\mu+r^{d-3}_{c}\sqrt{\Delta}_{c}}{r_{c}^{d-3}-\mu},\\
 \kappa&=&\frac{\Delta_c^{1/2}}{\sqrt{2}r_{c}(r_{c}^{d-3}-\mu)}\nonumber\\
     &\times& \sqrt{2a(d-3)(d-2) \mu
   (\sqrt{a-\Delta_c})
   r_c^{d-5}+\left(d^2-5d+2\right)
   \mu r_c^{d-3}+2 r_c^{2 d-6}-(d-4)(d-1)\mu^2}.\nonumber\\
\end{eqnarray}
Finally, in the small $a$ limit, we can expand the spacing of the entropy spectrum as
\begin{eqnarray}
 \Delta S&=&\frac{2\pi\mu r_{h}\sqrt{(1-r_{c}^{3-d}\mu)(8+2(d-1)(d-4)\mu r_{c}^{3-d})}}{r_{c}(2r_{h}^{d-3}+(d-5)\mu)}\hbar\nonumber\\
 &+&\frac{2\sqrt{2}\pi \mu^2 r_h r_c^{2(2-d)}
  \left((d-4)(d-1)\mu-(2+(d-5) d) r_c^{d-3}\right)}
    {\sqrt{(d-4)(d-1)\mu r_c^{3-d}+4} \left(2 r_h^{d-3}+(d-5)\mu\right)}a\hbar+\mathcal{O}(a^{2}).
\end{eqnarray}
The first term will go back to the Schwarzschild black hole case if $\mu$, $r_{c}$ and $r_{h}$ take the values as that of the Schwarzschild black hole. The second term looks like a corrected term from the spin $a$ of the black hole. The spacing for $d\geq 5$ is plotted against the dimensionless spin parameter $a/\mu^{1/(d-3)}$ in Figure \ref{PMP}. It shares the same behavior as the Kerr black hole. As $d$ increases, the spacing gets lower and lower, however the line becomes flat in high dimensions. In summary, we still obtain a spacing smaller than $2\pi\hbar$ in the small $a$ limit for $d\geq 5$.

\section{Summary}

In this paper, we show that the QNM frequencies of an asymptotically flat black hole in any dimension can be understood from the null geodesics. The physical view is thought that the QNM frequencies can be interpreted in terms of the massless particles trapped at the unstable circular null geodesics and slowly leaking out to infinity. The real part of the QNM frequencies is found to be the inverse of their impact parameter $\tilde{l}_{c}$ measured at the unstable circular null geodesics, and the imaginary part is $\kappa/\tilde{l}_{c}$.

Then using the new physical interpretation of QNM frequencies proposed by Maggiore, we calculate the quantum spectrum of the entropy for different black holes following the Hod's method. The result can be summarized as follows:

(i) The spacing of the entropy spectrum is dependent of the dimension $d$ of the spcetime. It decreases with $d$, and the largest spacing is $\Delta S\approx 2.1774\pi\hbar$ for $d=4$. When $d\geq 151$, the low bound of the spacing $\Delta S$ as suggested in \cite{BanerjeeVagenas} will be violated, and when $d\rightarrow\infty$, the spacing will vanish.

(ii) For a far from extremal black hole, the value of the spacing of the entropy spectrum is found to be larger than $2\pi\hbar$ for $d=4$ and smaller than $2\pi\hbar$ for $d\geq 5$, which is very different from these obtained from the previous work \cite{Vagenasjhep2008,Medvedcqg2008} (and references therein) by using the usual QNM frequencies.

In summary, since the spacing of the entropy spectrum in this paper is expressed in terms of the Hawking temperature and null geodesics of the black holes, this method is therefore justify to extend to other stationary black holes in non-Einstein gravity, and we conjecture that our result is universal.

\section*{Acknowledgement}
This work was supported by the National Natural Science Foundation of China (Grant No. 11205074 and Grant No. 11075065), and the Huo Ying-Dong Education Foundation of the Chinese Ministry of Education (Grant No. 121106).

\end{document}